\documentclass[a4paper,11pt]{article}
\usepackage{pos}
\usepackage{bm}
\usepackage{braket}
\usepackage{multicol}
\usepackage{float}

\title{Novel Bottomonium Results}
\author*[a]{Ben Page}
\author[a]{Chris Allton}
\author[b]{Seyong Kim}

\affiliation[a]{Department of Physics, Swansea University, Swansea SA2 8PP, United Kingdom}

\affiliation[b]{Department of Physics, Sejong University, Seoul 143-747, Korea}

\emailAdd{\{b.page.900727,c.r.allton\}@swansea.ac.uk}
\emailAdd{skim@sejong.ac.kr}

\abstract{
We present the latest results from the use of the Backus-Gilbert method for reconstructing the spectra of NRQCD bottomonium mesons using anisotropic FASTSUM ensembles at non-zero temperature. We focus in particular on results from the $\eta_b$, $\Upsilon$, $\chi_{b1}$ and $h_b$ generated from Tikhonov-regularized Backus-Gilbert coefficient sets. We extend previous work on the Laplace shifting theorem as a means of resolution improvement and present new results from its use. We conclude with a discussion of the limitations of the improvement routine and elucidate a connection with Parisi-Lepage statistical scaling.
}

\FullConference{%
 The 39th International Symposium on Lattice Field Theory,\\
  8th-13th August, 2022,\\
  Rheinische Friedrich-Wilhelms-Universit\"at Bonn, Bonn, Germany

}

\begin{document}
\maketitle

\section{Introduction}

Bottomonium has previously been studied as a means of estimating the properties of the quark-gluon plasma as produced in relativistic heavy-ion collisions. \cite{aarts_bottomonium_2010,aarts_what_2011,aarts_bottomonium_2014}. The problem of reconstructing the spectrum of bottomonium states at non-zero temperature is an ill-posed one due to the presence of finite uncertainties in data collected from lattice QCD simulations. This work represents one of the latest in a collection of studies performed by the {\sc Fastsum} Collaboration focussing on bottomonium states simulated using anisotropic lattices at nonzero temperature. In the following, we build upon the results of previous work \cite{page_spectral_2021} and present a discussion on the use of the Backus-Gilbert method for reconstructing bottomonium spectra, showing how the Laplacian nature of the reconstruction formula my be exploited to give improved resolution.

\section{Lattice details}

Non-relativistic QCD (NRQCD) is an effective field theory for heavy quarkonia, which approximates fully relativistic QCD by expanding the Lagrangian in powers of the heavy quark velocity \cite{lepage_simulating_1992}. One of the principal benefits of the NRQCD formulation is that the calculation of the time evolution of the heavy quarks reduces to an initial-value problem, as the heavy quarks and antiquarks decouple in the non-relativistic regime. This decoupling effect turns the spectral representation of the Euclidean correlator $G(\tau)$ into a Laplace transformation of the spectral density function $\rho(\omega)$:
\begin{equation}
    G(\tau;T)=\int_{\omega_\text{min}}^{\omega_\text{max}}
    \frac{d\omega}{2\pi}
    K(\tau,\omega)\rho(\omega;T)
    \label{eq:spectral_repr}
\end{equation}
where $K(\tau,\omega)=e^{-\omega\tau}$ is the temperature independent kernel function and $T=(a_\tau N_\tau)^{-1}$ is the lattice temperature as a function of the temporal extent $N_\tau$. In order to relate NRQCD energies to physical energies, the reconstruction window $\omega\in[\omega_\text{min},\omega_\text{max}]$ must be additively renormalised by the NRQCD energy shift, $\Delta E = 7.46$GeV. 

We make use of {\sc Fastsum}'s Generation 2L ensembles, generated using anisotropic lattices ($\xi = a_s/a_\tau \sim 3.5$) with 2+1 flavour, clover-improved Wilson fermions using a  physical $s$ quark and lighter, degenerate $u$ and $d$ quarks\cite{aarts_properties_2020}. The spatial extent of the lattice $N_s=32$ and the temporal extent along with the corresponding lattice temperatures are detailed in Table \ref{tab:Ntau_correlators}.

\begin{table}[h]
    \centering
 \begin{tabular}{|c || c | c | c | c | c || c | c | c | c | c | c |} 
 \hline
  $N_{\tau}$ & 128 & 64 & 56 & 48 & 40 & 36 & 32 & 28 & 24 & 20 \\ [0.5ex] 
 \hline
  $T=1/(a_\tau N_\tau)$ [MeV] & 47 & 95 & 109 & 127 & 152 & 169 & 190  & 217 & 253 & 304\\
 \hline 
 $N_{\rm cfg}$ & 1024\! & 1041\! & 1042\! & 1123\! & 1102\! & 1119\! & 1090\!  & 1031\! & 1016\! & 1030\!\\
 \hline
\end{tabular}
 \caption{Temporal extent, corresponding lattice temperature in MeV and number of configurations for the {\sc Fastsum} Generation 2L ensembles. The double vertical line mid-table represents our value of $T_{\rm pc}$\cite{aarts_properties_2020}.}
    \label{tab:Ntau_correlators}
\end{table}

\section{The Backus-Gilbert Method}

The Backus-Gilbert method \cite{backus_resolving_1968} is a reconstruction technique which extracts regularised solutions from the ill-posed inverse problem by imposing constraints on the stability of its predictions under a change of input. Since $G(\tau)$ is only known to at most $\mathcal{O}(100)$ but $\rho(\omega)$ is continuous ($\mathcal{O}(1000+)$ points), there are theoretically an infinite number of possible spectra $\rho$ which produce the correct $G(\tau)$ within numerical errors. Backus-Gilbert attempts to estimate a solution of Eq. \ref{eq:spectral_repr}, denoted $\hat{\rho}$, on a point-by-point basis by constructing averaging functions $A(\omega,\omega_0)$ centred about some point $\omega_0$ generated using the data kernel $K(\tau,\omega)$:
\begin{equation}
    \hat{\rho}({\omega_0}) = \int^{\omega_\text{max}}_{\omega_\text{min}} A(\omega,\omega_0) \rho(\omega)~d\omega
    \label{eq:bg_estimate_integral}
\end{equation}
with $A(\omega,\omega_0)=\sum_\tau c_\tau(\omega_0)K(\tau,\omega)$. In the limit $A(\omega,\omega_0) \longrightarrow \delta(\omega-\omega_0)$, we obtain a perfect reconstrucion of the target spectrum $\rho$. It is easily seen that plugging Eq.~\ref{eq:spectral_repr} into
\begin{equation}
    \hat{\rho}(\omega_0) = \sum_\tau c_\tau(\omega_0) G(\tau).
    \label{eq:bg_estimate_sum}
\end{equation}
gives Eq.~\ref{eq:bg_estimate_integral}. The coefficients $c_\tau(\omega_0)$ control the shape of $A(\omega,\omega_0)$ and are found by minimising the cost function $J(\omega_0)$ representing the least-squares distance between the averaging function and the delta function at $\omega_0$\cite{oldenburg_introduction_1984}:
\begin{equation}
    J(\omega_0) = \int_{\omega_\text{min}}^{\omega_\text{max}} \left[A(\omega,\omega_0)-\delta(\omega-\omega_0)\right]^2~d\omega.
    \label{eq:dirichlet criterion}
\end{equation}
Setting $\partial_{c_\tau} J(\omega_0) = 0$ reduces the problem to an inversion of a matrix-vector product:
\begin{equation}
    \mathcal{K}_{\tau\tau'} \cdot c_{\tau'}(\omega_0)= K(\omega_0,\tau),\qquad \text{where}\quad \mathcal{K}_{\tau\tau'}= \int_{\omega_\text{min}}^{\omega_\text{max}} K(\tau, \omega) K(\tau', \omega) d\omega.
    \label{eq:dirichlet matrix}
\end{equation}

The kernel width matrix $\mathcal{K}$ is near-singular (worsened by the exponential nature of the kernel) and so must be treated by a conditioning routine before inversion. One such conditioning routine is the addition of a small constant term to the diagonal entries in a Tikhonov-like fashion\cite{tikhonov_stability_1943}:
\begin{equation}
    \mathcal{K}(\alpha) = \mathcal{K} + \alpha I
    \label{eq:tikhonov_conditioning}
\end{equation}
where $\alpha$ is a parameter which controls the strength of the regularisation. The benefit of such scalar conditioning over other methods is that the coefficients $c_\tau$ are constructed without prior knowledge of $G(\tau)$, enabling their use in the reconstruction of any spectra obeying Eq.~\ref{eq:spectral_repr}, regardless of choices of quantum numbers, see \cite{page_spectral_2021}.

\section{Improvement via the Laplace Shift Transform}
\label{sec:laplace_shift}

It has been previously shown \cite{rothkopf_improved_2012} that, for the case of the maximum entropy method (MEM), there exists a relationship between the choice of $\omega_\text{min}$ and the resolving power of the method. This effect also occurs in the Backus-Gilbert method which shares a similar basis-function mechanism of reconstruction as the MEM. The application of a Laplace shift transform:
\begin{equation}
  G'(\tau) = e^{\Delta\cdot\tau}G(\tau)  
  \overset{\mathcal{L}}{\Longrightarrow}
  \rho'(\omega) = \rho(\omega+\Delta),
    \label{eq:laplace shift rule}
\end{equation}
where $\Delta>0$ shifts the spectral features closer to $\omega_\text{min}$ where the resolving power of the method is improved \cite{page_spectral_2021} offering improved predictions for mass and width estimates. Building upon this feature, we have opted to combine $\omega_\text{min}$ and the Laplace shift transform parameter $\Delta$ into a single parameter $\widetilde{\Delta}=\omega_\text{min}+\Delta$.

\section{Systematic analysis: Removing $\widetilde{\Delta}$ and $\alpha$ dependence}
\label{sec:systematic_errors}

\begin{figure}[H]
\includegraphics[width=0.49\linewidth]{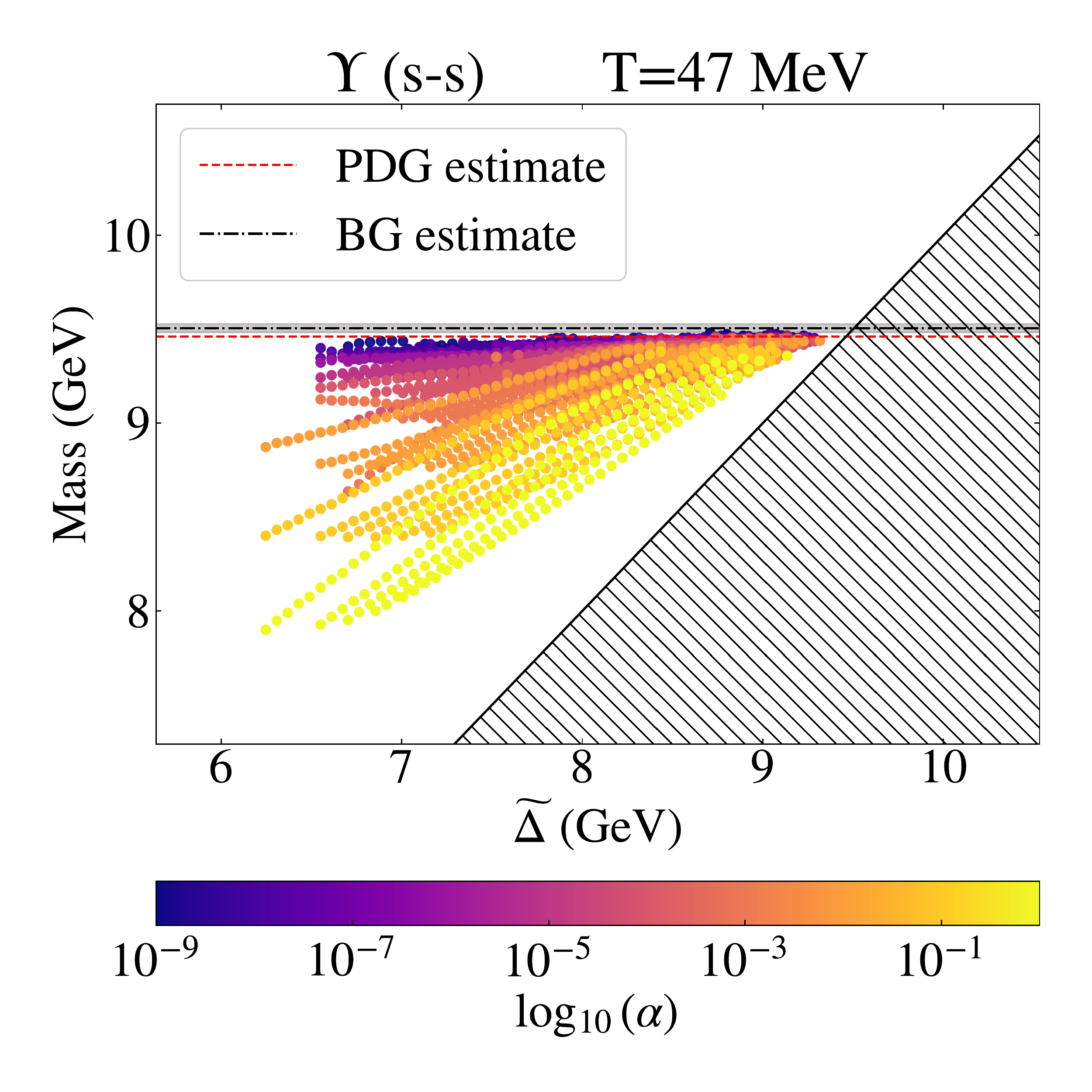}
\includegraphics[width=0.49\linewidth]{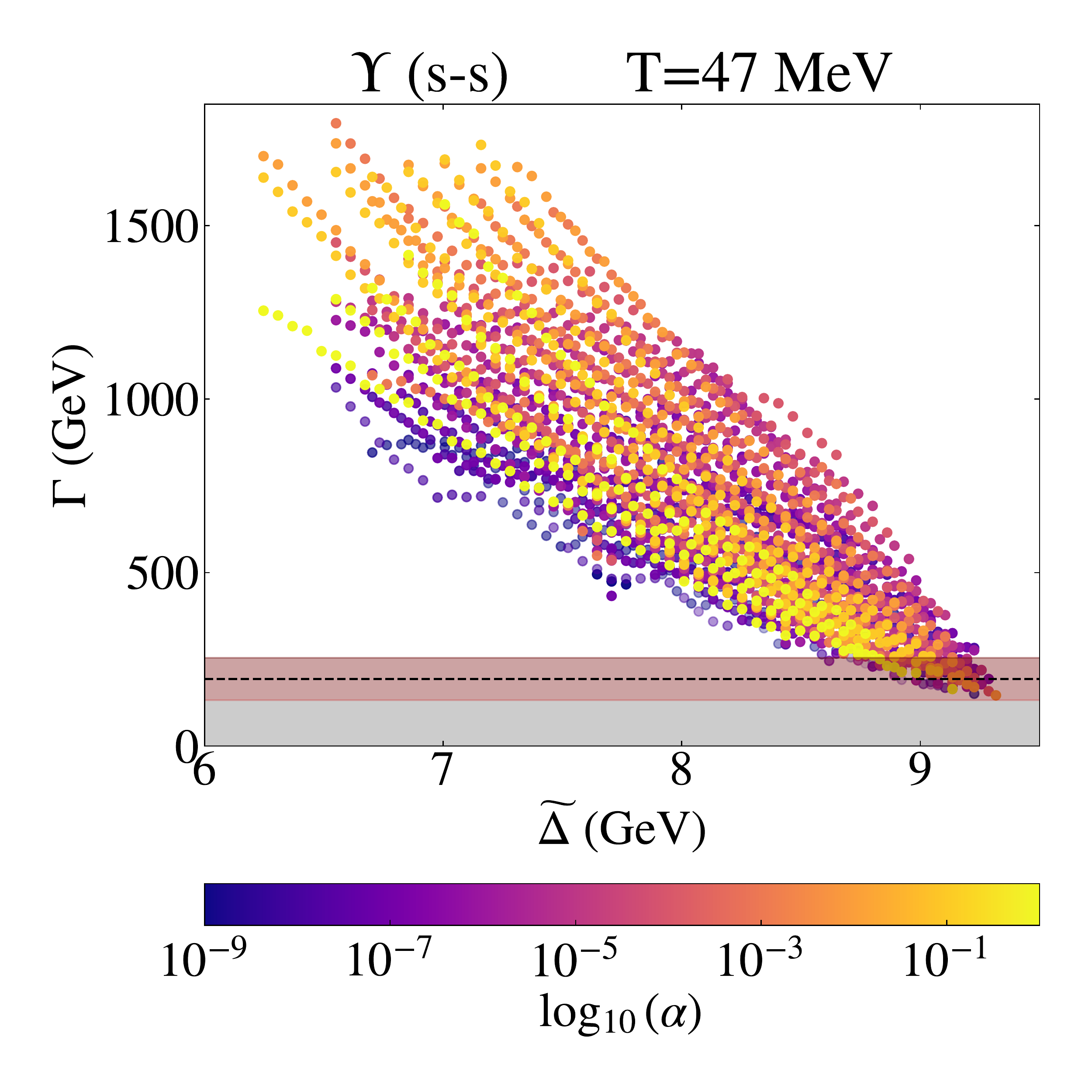}
\caption{\emph{Left:} Plot of the ground state mass versus the shift parameter $\widetilde{\Delta}$ for the (smeared) $\Upsilon$ meson over a range of $\alpha$ values. The hatched region represents the maximum possible shift, beyond which the ground state feature falls outside of the sampling window. The red dashed line is the PDG estimate \cite{particle_data_group_review_2022} and the black dashed line represents our best estimate after extrapolating the $\widetilde\Delta$ and $\alpha$ hyper-parameters. \emph{Right}: Plot of our estimate of the upper bound for the ground state FWHM width versus $\widetilde\Delta$ for the $\Upsilon$ meson over a range of $\alpha$ values. The black dashed line is our best estimate of the upper bound and the red band denotes its uncertainty. The temperature used in these plots is $T=47$ MeV.}
\label{fig:breadcrumb_example}
\end{figure}

There is a systematic dependence of the ground state mass $M$ and width $\Gamma$ on the value of $\widetilde{\Delta}$ and $\alpha$ which must be removed during analysis. This is illustrated in Fig.~\ref{fig:breadcrumb_example} where the mass (Left) and width (Right) are shown as a function of $\widetilde{\Delta}$ for various $\alpha$ values.

As can be seen, for fixed $\alpha$, the mass increases and width decreases monotonically with $\widetilde{\Delta}$ due to the resolution improvement. 
We also note that in the mass case, this slope approaches zero as $\alpha\rightarrow0$, indicating that the $\widetilde{\Delta}$ dependence falls away.
Finally, we note that the maximum theoretical shift occurs when $\widetilde{\Delta}$ equals the mass. This is indicated by the boundary of the hatched region in Fig.~\ref{fig:breadcrumb_example} (Left)). 
We therefore obtain our mass estimate by firstly performing a linear extrapolation of the mass, at fixed $\alpha$, to the boundary of the hatched region, obtaining $M(\alpha)$.
The $\alpha$ dependence is then removed by noting empirically that $M(\alpha)$ becomes independent of $\alpha$, for small $\alpha$, and so can be fit with a constant. A bootstrap analysis produces the error estimate.

The ground state width is determined by a similar procedure. First a linear fit in $\widetilde{\Delta}$ is performed, at fixed $\alpha$, and extrapolated to the boundary region.
The $\alpha$ dependence is then removed in the same manner as in the mass case.
However we note that, due to the finite resolving power of the Backus Gilbert method, the widths obtained should be considered {\em upper bounds} on the physical width of the state.

The value of $\omega_\text{min}$ is held fixed in this analysis (at a value of $\omega_\text{min}=-0.1a_\tau^{-1}\approx6.8$GeV).
We have chosen to include all possible Euclidean times in our analysis, i.e. $0\leq\tau<N_\tau$, because the resolving power of Backus Gilbert method is greatest for the largest time windows.

\section{Results for the bottomonium sector: $\eta_b$, $\Upsilon$, $\chi_{b1}$ and $h_b$}

\begin{figure}[H]
    \centering
    \includegraphics[width=0.495\linewidth]{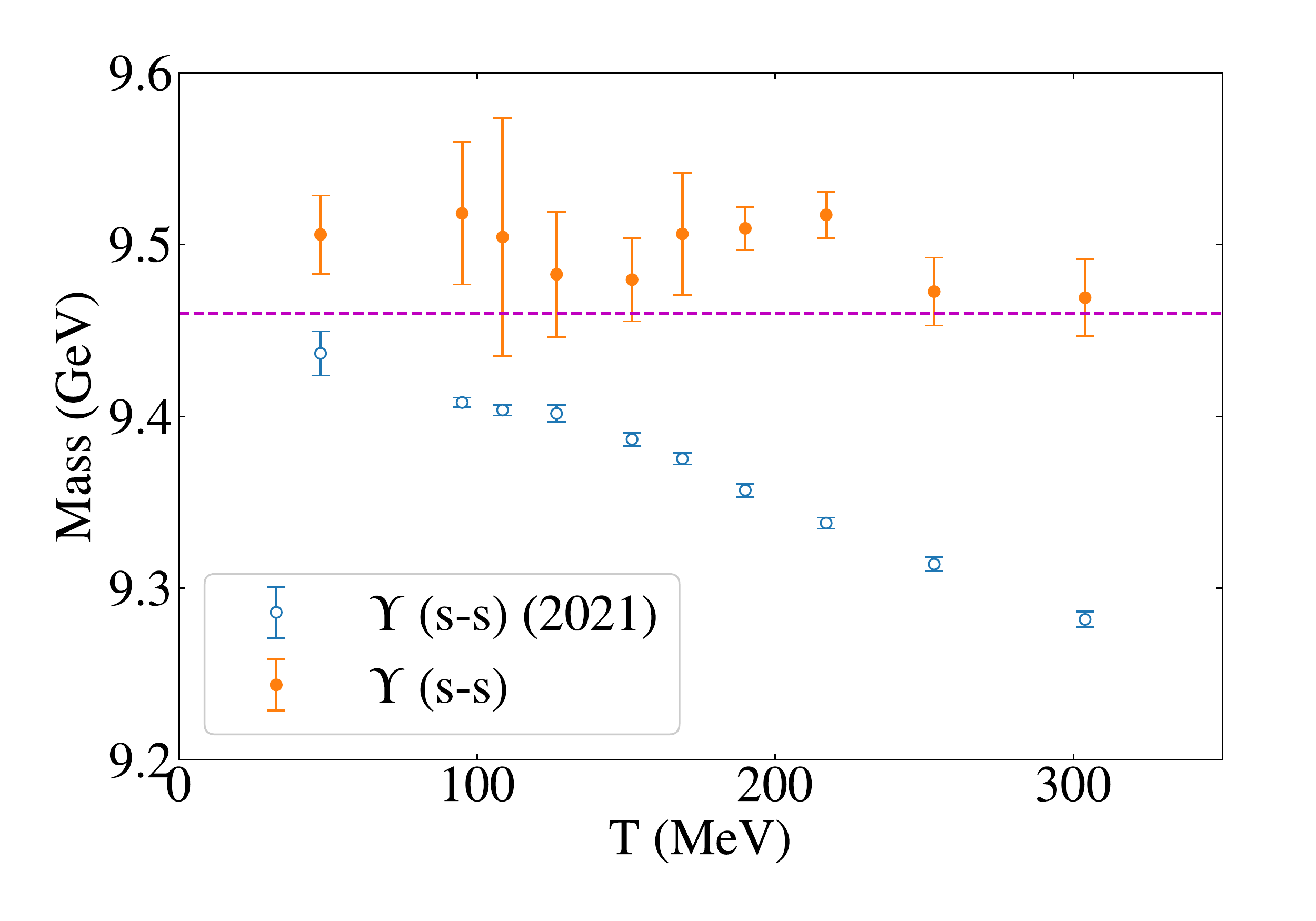}
    \includegraphics[width=0.495\linewidth]{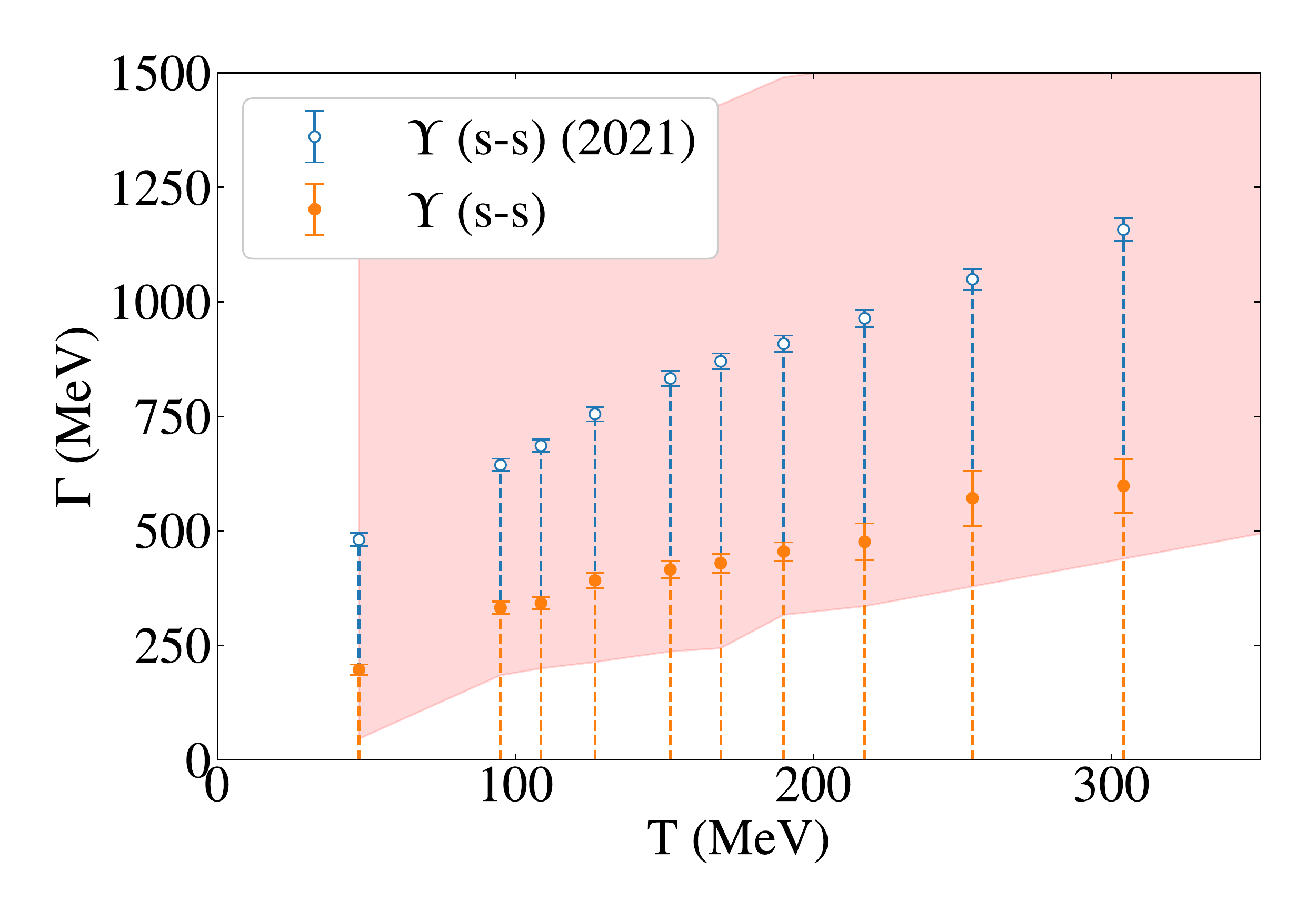}
    \caption{Comparison between new results from this work and those from \cite{page_spectral_2021}, labelled (2021), for the mass (\emph{left}) and upper bound on the width (\emph{right}). The magenta line on the left pane is the PDG estimate for the $\Upsilon$ mass \cite{particle_data_group_review_2022}. The red shaded band on the right pane indicates our estimate of the maximum resolving power of the new method.}
    \label{fig:2021_2022_comparison}
\end{figure}

\begin{figure}[h]
    \centering
    \includegraphics[width=0.9\linewidth]{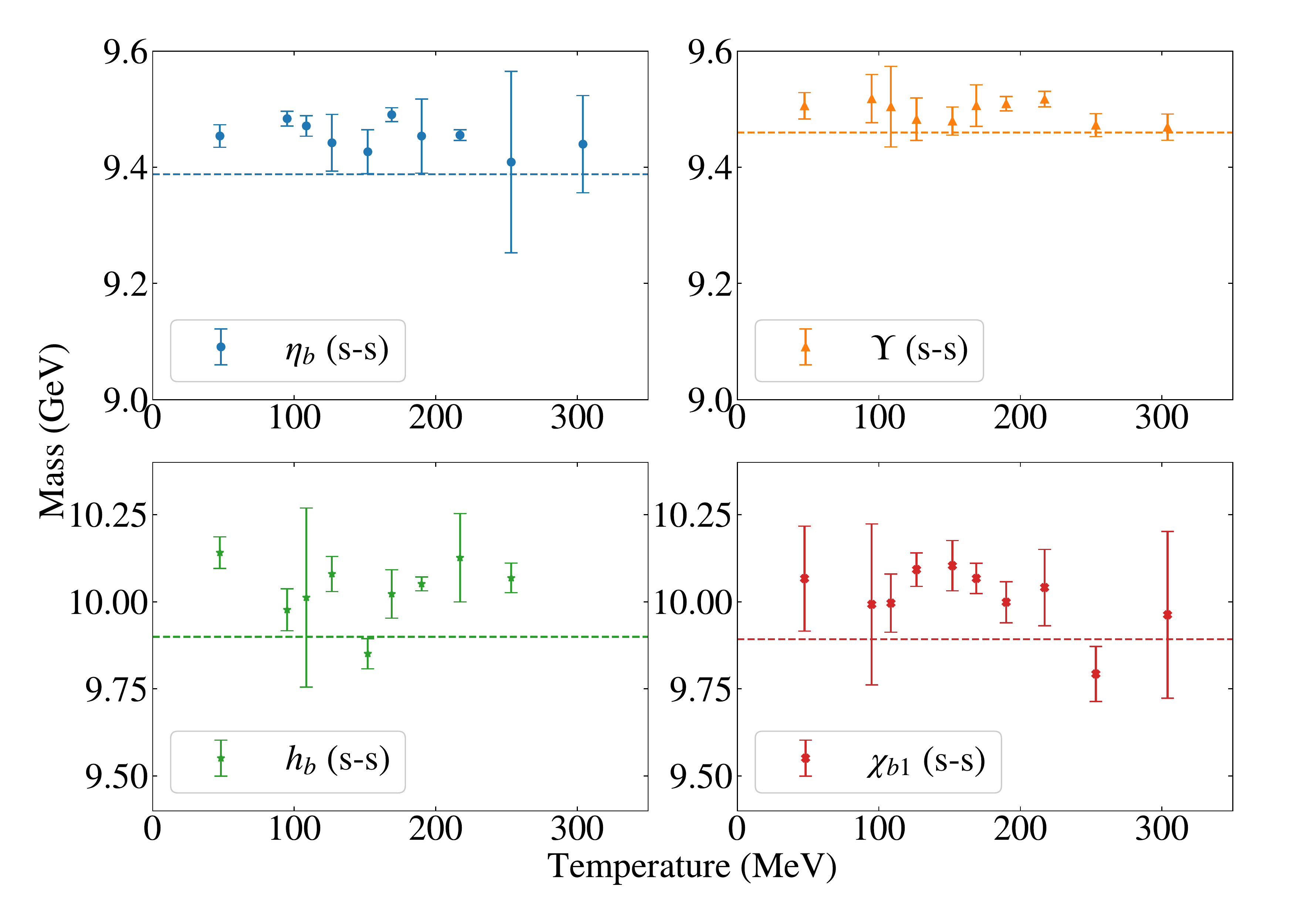}
    \caption{Plots showing the estimate of the mass versus lattice temperature for select bottomonium states. The horizontal dashed line represents the PDG estimate for the given state \cite{particle_data_group_review_2022}.}
    \label{fig:2022_all_mu}
\end{figure}

\begin{figure}[H]
    \centering
    \includegraphics[width=0.9\linewidth]{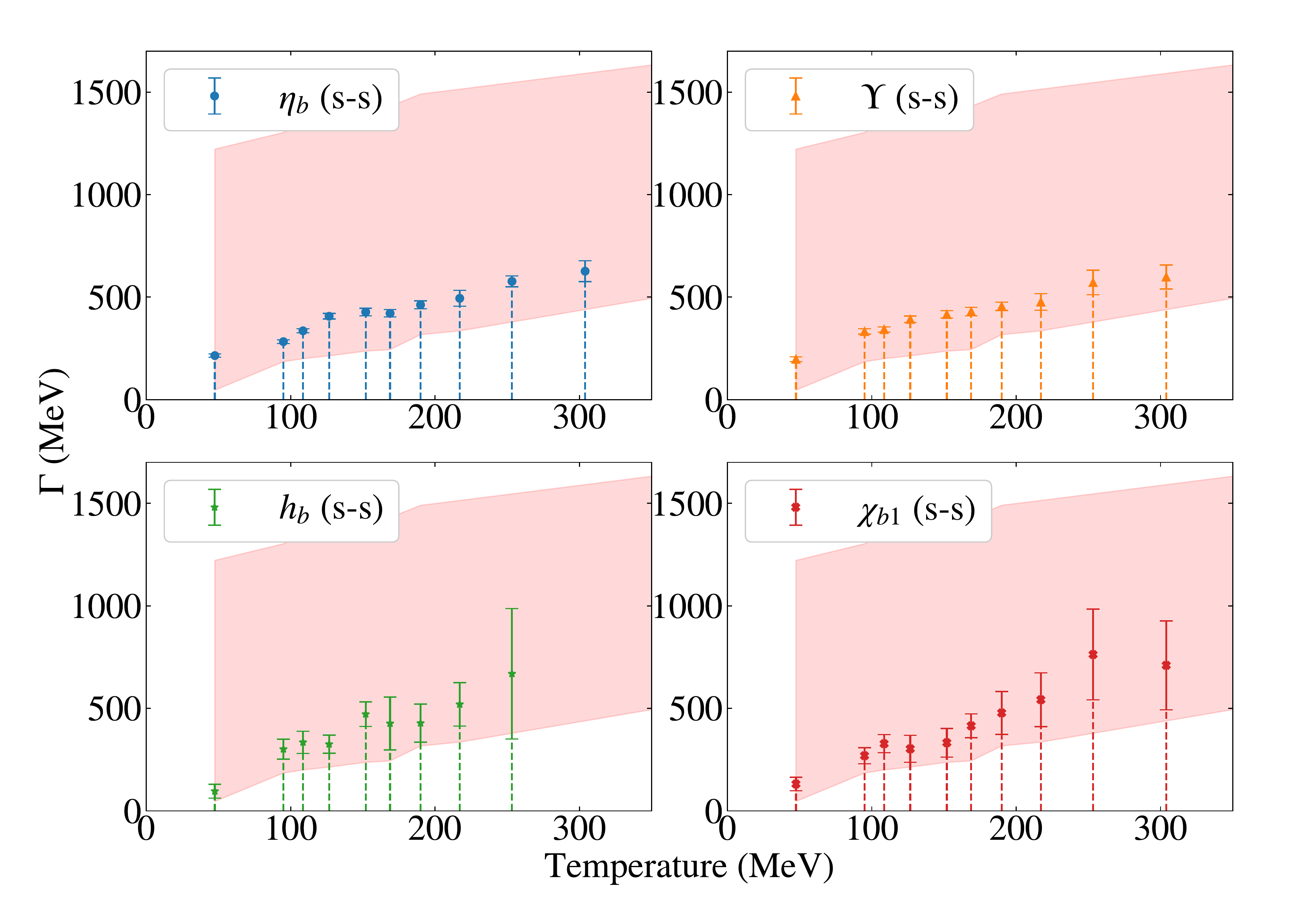}
    \caption{Plots showing the estimate of the upper bound on the width versus lattice temperature for select bottomonium states. The shaded region represents our best estimate of the resolving power of the method.}
    \label{fig:2022_all_fwhm}
\end{figure}

We have improved our previous results \cite{page_spectral_2021}) by including the Laplace shifting (see \S\ref{sec:laplace_shift}) and an improved fitting analysis (see \S\ref{sec:systematic_errors}).
In Fig.~\ref{fig:2021_2022_comparison}, we display our results for the $\Upsilon$ mass and width as a function of temperature, including the results from our earlier analysis.
We have also extended our previous work by including the $\eta_b$, $\chi_{b1}$ and $h_b$ states, with results shown in Figs.~\ref{fig:2022_all_mu} and \ref{fig:2022_all_fwhm}.
For this analysis, we restricted ourselves to data generated using smeared quark sources which have improved overlap with the ground state over local sources. We also conducted a more rigorous study of the dependence of the mass and width on the parameters $\alpha$ and $\widetilde{\Delta}$ in an attempt to measure their contribution to the systematic error.

In particular, in the case of the $\Upsilon$ mass, we wish to highlight in Fig.\ref{fig:2021_2022_comparison} the comparison between the results of this work and the estimates presented in \cite{page_spectral_2021} which appear to be contaminated by systematics of the method, as the dependence of the the mass and width on $\alpha$ was not fully accounted for. We also note that below the pseudocritical temperature ($T_{\rm pc}=162~$MeV for our Gen-2L ensembles\cite{aarts_properties_2020,aarts_open_2022}), the experimental widths for the $\Upsilon$ and $\eta_b$ are $54.02\pm1.25$~keV and $10${\raisebox{0.25ex}{\small$\substack{+4 \\ -5}$}}~MeV respectively\cite{particle_data_group_review_2022}, over an order of magnitude smaller than even the minimum resolvable width for our method (see Fig.~\ref{fig:2022_all_fwhm}). We once again reiterate that our presented values for the ground state width represent an upper bound on the true value, and we leave an investigation into the effect of changing the Euclidean time extent and the lower bound $\omega_\text{min}$ to a future study.
\bigskip
\bigskip
\bigskip
\bigskip


\begin{figure}[h]
    \centering
    \includegraphics[width=0.8\linewidth]{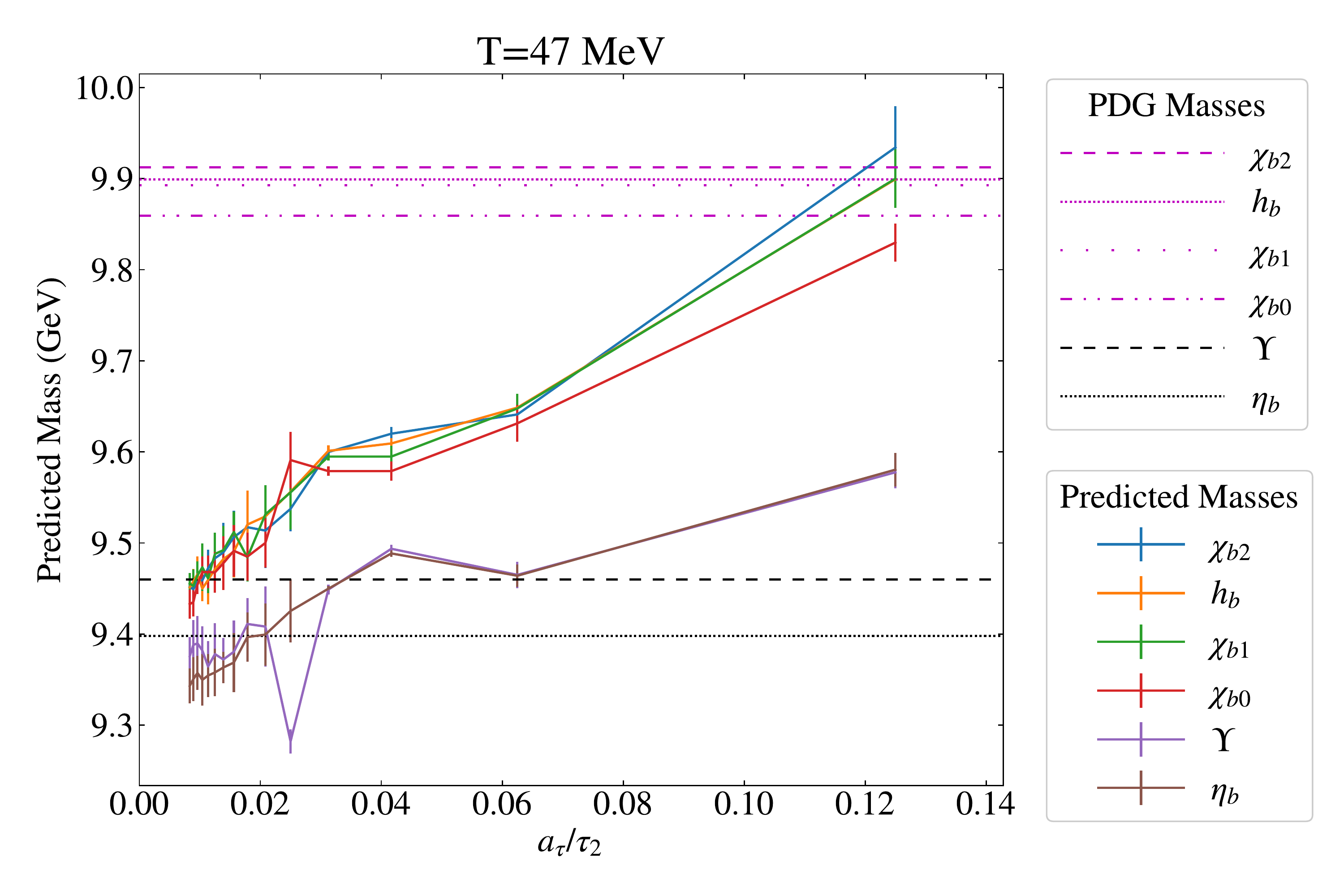}
    \caption{Value of the energy shift $\Delta_\text{sing}$ (i.e. the predicted mass) which gives the most singular shifted covariance matrix (see Eq. \ref{eq:cov}) for a variety of bottomonium channels as a function of $1/\tau_2$. The covariance matrices are defined over the time interval $0\leq\tau<\tau_2$, and therefore the best results are obtained as $\tau_2 \rightarrow \infty$. The lattice temperature is 47 MeV. The predicted masses each meson tends toward the experimental estimate for the pseudoscalar mass. Experimental values for the meson masses are shown as horizontal lines \cite{particle_data_group_review_2022}.}
    \label{fig:smeared_cov}
\end{figure}

\section{Connection with Parisi-Lepage Statistical Scaling}

To estimate the error in the resulting reconstruction, the uncertainty in the Euclidean correlator $\Delta G(\tau)$ must be combined with the Backus Gilbert coefficients $c_\tau$. The uncertainty corresponding to Eq.~\ref{eq:bg_estimate_sum} is simply
\begin{equation}
    \Delta\hat{\rho}^2 = \sum_{\tau,\tau'} c_{\tau} \text{Cov}[G]_{\tau,\tau'} c_{\tau'}
    \label{eq:spectral_error}
\end{equation}
where $\text{Cov}[G]$ is the covariance in $G(\tau)$. Under the Laplace transformation outlined in Eq.~\ref{eq:laplace shift rule}, the covariance matrix transforms as
\begin{equation}
    \text{Cov}[G;\Delta]_{\tau,\tau'} = e^{\Delta\cdot\tau}\text{Cov}[G]_{\tau\tau'}e^{\Delta\cdot\tau'}
\label{eq:cov}
\end{equation}
which in turn influences the spectral error $\Delta\hat\rho$. This effect can be probed by measuring the condition number of the resulting matrix, defined by
\begin{equation}
    \kappa\left(\text{Cov}[G;\Delta]\right) = \dfrac{\sigma_\text{max}}{\sigma_\text{min}}
\end{equation}
where $\sigma$ are the singular values of the matrix. 

It is interesting to study $\kappa$ as a function of $\Delta$ and determine the value, $\Delta_\text{sing}$ when $\text{Cov}[G;\Delta]$ becomes singular. One may imagine that $\Delta_\text{sing}$ is the ground state mass of the $G(\tau)$ channel. However, as pointed out by Parisi \cite{parisi_strategy_1984} and elucidated further by Lepage \cite{lepage_analysis_1989}, the covariance matrix has a special physical significance. It can be expressed as a correlation function of the {\em square} of the interpolating operators of the original correlation function, $G(\tau)$. Analysing this further, one finds that the lightest state which contributes to the covariance matrix is the pseudoscalar, no matter what state was being probed by $G(\tau)$. This therefore implies that $\text{Cov}[G;\Delta]$ becomes singular when $\Delta=\Delta_\text{sing}$ is the pseudoscalar mass (i.e. the $\eta_b$ mass in our case) independent of the channel.

We illustrate this in Fig. \ref{fig:smeared_cov} where $\Delta_\text{sing}$ is plotted for a variety of channels. The covariance matrix was defined over the time interval $0\leq\tau<\tau_2$ meaning that the large time limit (where the ground state dominates) is obtained as $\tau_2\rightarrow\infty$. As can be seen, in this limit we recover the $\eta_b$ (i.e. pseudoscalar) mass, thereby confirming the prediction of \cite{parisi_strategy_1984,lepage_analysis_1989}.



\section{Summary}
We have presented results for the ground state mass and an estimate for the upper bound on the width for several bottomonium states using smeared quark sources, showing improved resolution compared to our previous results. We have demonstrated the ability of the Laplace shift to naively increase the resolving power of the method, but show that is still insufficient to resolve the ground state widths of the system. The effect of the Laplace shift transform on the covariance matrix of the Euclidean correlation function was also studied, where the condition number of the matrix was found to confirm Parisi-Lepage statistical scaling in the long-time limit.

\acknowledgments 

This work is supported by STFC grant ST/T000813/1.
SK is supported by the National Research Foundation of Korea under grant NRF-2021R1A2C1092701 and grant NRF-2021K1A3A1A16096820, funded by the Korean government (MEST).
BP has been supported by a Swansea University Research Excellence Scholarship (SURES).
This work used the DiRAC Extreme Scaling service at the University of Edinburgh, operated by the Edinburgh Parallel Computing Centre and the DiRAC Data Intensive service operated by the University of Leicester IT Services on behalf of the STFC DiRAC HPC Facility (www.dirac.ac.uk). This equipment was funded by BEIS capital funding via STFC capital grants ST/R00238X/1, ST/K000373/1 and ST/R002363/1 and STFC DiRAC Operations grants ST/R001006/1 and ST/R001014/1. DiRAC is part of the UK National e-Infrastructure.
This work was performed using PRACE resources at Cineca (Italy), CEA (France) and Stuttgart (Germany) via grants 2015133079, 2018194714, 2019214714 and 2020214714.
We acknowledge the support of the Swansea Academy for Advanced Computing, the Supercomputing Wales project, which is part-funded by the European Regional Development Fund (ERDF) via Welsh Government, and the University of Southern Denmark and ICHEC, Ireland for use of computing facilities.
We are grateful to the Hadron Spectrum Collaboration for the use of their zero temperature ensemble.

\bibliographystyle{JHEP}
\bibliography{Lattice2022}

\end{document}